\documentclass[aps,prd,reprint,superscriptaddress,preprintnumbers,nofootinbib]{revtex4-1}

\usepackage{amsmath}
\usepackage{amsfonts} 
\usepackage{amssymb}
\usepackage{graphicx}
\usepackage{hyperref}
\usepackage{tensor}
\usepackage{siunitx}

\begin{document}

\title{\Large A Spin-2 Conjecture on the Swampland}
\preprint{MPP-2018-276}
\preprint{LMU-ASC-76/18}

\author{\large Daniel Klaewer}
\affiliation{Max-Planck-Institut f\"ur Physik (Werner-Heisenberg-Institut),\\
             F\"ohringer Ring 6,
             80805, M\"unchen, Germany}
             \author{\large Dieter L\"ust}
\affiliation{Arnold-Sommerfeld-Center for Theoretical Physics, Ludwig-Maximilians-Universit\"at, 80333 M\"unchen, Germany}
\affiliation{Max-Planck-Institut f\"ur Physik (Werner-Heisenberg-Institut),\\
             F\"ohringer Ring 6,
             80805, M\"unchen, Germany}
\author{\large Eran Palti}
\affiliation{Max-Planck-Institut f\"ur Physik (Werner-Heisenberg-Institut),\\
             F\"ohringer Ring 6,
             80805, M\"unchen, Germany}

\begin{abstract}
\vspace{0.5cm}
We consider effective theories with massive fields that have spins larger than or equal to two. We conjecture a universal cutoff scale on any such theory that depends on the lightest mass of such fields. This cutoff corresponds to the mass scale of an infinite tower of states, signalling the breakdown of the effective theory. The cutoff can be understood as the Weak Gravity Conjecture applied to the St\"uckelberg gauge field in the mass term of the high spin fields. A strong version of our conjecture applies even if the graviton itself is massive, so to massive gravity. We provide further evidence for the conjecture from string theory.
\vspace{1cm}
\end{abstract}

\maketitle

\section{A Spin-2 Swampland Conjecture}
\label{sec:s2c}

There are many effective field theories that appear completely consistent in the infrared, but cannot be completed to quantum gravity in the ultraviolet. Such theories are termed to be in the Swampland \cite{Vafa:2005ui}. 
One class of theories that has received considerable attention are theories which contain massive  fields with spin larger than or equal to two. 
Perhaps the most studied of these are so-called massive or bi-gravity theories. See, for example, reviews on massive gravity \cite{deRham:2014zqa} and bi-gravity \cite{Schmidt-May:2015vnx} theories. In this note we propose a Swampland criterium on such theories in the form of a universal quantum gravity bound on their cutoff. 

We will focus on the most interesting case of spin-2. We consider an action with Einstein gravity and an additional massive spin-2 field $w_{\mu\nu}$ of mass $m$:
\begin{eqnarray}
S_G &=& \int d^4x \; \sqrt{-G} \left[ M^2_p R\left(G\right)  - \frac14 w^{\mu\nu} L_{\mu\nu}^{\rho\sigma} w_{\rho \sigma} \right. \nonumber \\ 
& &\left.-\frac18 m^2 \left( w_{\mu\nu}w^{\mu\nu} - w^2 \right) +... \right] \;. 
\label{ehgen}
\end{eqnarray}
Here $R\left(G\right)$ is the Ricci scalar constructed from $G$, $\sqrt{-G}$ is the associated determinant, and $M_p$ is the Planck mass. We will consider for simplicity at this stage a flat space background $G_{\mu\nu}=\eta_{\mu\nu}$, but will generalise it later. $L_{\mu\nu}^{\rho\sigma}$ is then the Lichnerowicz operator, which in flat space takes the form
\begin{eqnarray}
L_{\mu\nu}^{\rho\sigma} w_{\rho\sigma} &=& -\frac12 \left[ \Box w_{\mu\nu} - 2\partial_{(\mu}\partial_{\alpha} w^{\alpha}_{\nu)} + \partial_{\mu}\partial_{\nu} w \right.\nonumber \\ 
& &\left.- \eta_{\mu\nu} \left( \Box w - \partial_{\alpha} \partial_{\beta} w^{\alpha\beta}\right) \right] \;.
\end{eqnarray}
The last term in (\ref{ehgen}) is the Fierz-Pauli mass, and $w \equiv \eta^{\mu\nu}w_{\mu\nu}$. We have linearised the action in $w_{\mu\nu}$, but will discuss in section \ref{sec:hd} a full non-linear completion in the form of bi-gravity. 

A symmetric tensor has 10 degrees of freedom.  And as is well-known,
a massless spin-2 field propagates 2 degrees of freedom, with 8 modes removed using local diffeomorphisms. 
On the other hand, a massive spin-2 field propagates 5 degrees of freedom, and these can be thought of as being composed of the 2 degrees of freedom of a massless spin-2 field, 2 degrees of freedom from a massless gauge vector, and 1 scalar degree of freedom. Explicitly, we can decompose $w_{\mu\nu}$ in terms of these degrees of freedom as (see, for example, \cite{Dvali:2006su})
\begin{equation} 
\label{decompo}
w_{\mu\nu} = h_{\mu\nu} + 2 \partial_{(\mu}\chi_{\nu)} + \Pi^L_{\mu\nu} \pi \;.
\end{equation} 
Here, $h_{\mu\nu}$ is the helicity-2 mode, $\chi_{\mu}$ the helicity-1 mode, and $\pi$ the scalar longitudinal mode, with $\Pi^L_{\mu\nu}$ being the associated projector. 
The mass term then contains the kinetic term for $\chi_{\mu}$
\begin{equation}
\label{lfpfs}
{\cal L}_{\mathrm{FP}} \supset -\frac{1}{8}m^2 F_{\mu\nu} F^{\mu\nu} \;, 
\end{equation}
where $F_{\mu\nu} \equiv \partial_{\mu} \chi_{\nu} - \partial_{\nu}\chi_{\mu}$.

The helicity-1 mode will couple to some matter current $J^{\mu}$ through an interaction of the form
\begin{equation}
\label{h1int}
{\cal L}_{\mathrm{int}} =  \frac{m^2}{M_w} \chi_{\mu} J^{\mu} \;,
\end{equation}
where we have introduced an interaction mass scale $M_w$. To understand this behaviour one notes that the spin-2 mode $w_{\mu\nu}$ will couple to a tensor $T_w^{\mu\nu}$. This tensor is conserved if $w_{\mu\nu}$ is massless by gauge symmetry, the mass term breaks this symmetry leading to non-conservation of $T_w^{\mu\nu}$. So, after integration by parts, one finds an interaction $\chi_{\mu} \left( \partial_{\nu} T_w^{\mu\nu} \right)$ which must be proportional to $m^2$. Completing the dimensions by an interaction scale $M_w$ associated to the spin-2 mode, so the analogue of the Planck mass for the graviton, gives the form (\ref{h1int}).

Our conjecture is then formulated as 

\medskip
\noindent
{\bf Spin-2 Conjecture:} {\it An effective theory with a self-interacting massless spin-2 field (Einstein gravity) and which additionally contains a field of spin 2, mass $m$, and associated interaction mass scale $M_w$, has a universal cutoff $\Lambda_{m}$ with}
\begin{equation}
\label{s2c}
\Lambda_{m} \sim \frac{m \;M_p}{M_w} \;.
\end{equation}
{\it This cutoff is associated with the mass scale of an infinite tower of states.}
 \medskip

This means that spin-two fields cannot have arbitrarily low mass without lowering the cut-off scale $\Lambda_{m} $ accordingly.
Note that in the case of multiple massive spin-2 fields, the conjecture should be applied to the one which leads to the lowest-energy cutoff $\Lambda_{m} $. The conjecture should also be taken to apply only if the spin-2 field retains its fundamental nature, rather than splitting into constituents, up to the mass scale $\Lambda_{m} $. 
Since it is natural to assume that $M_w \leq M_p$, one obtains that the cut-off scale obeys $\Lambda_{m}\geq  m$.

We will present evidence for this conjecture from string theory. From a more general perspective, the conjecture can be thought of as a quantum gravity obstruction to taking the continuous limit leading to two massless spin-2 fields. However, the most general and quantitative evidence comes from the Weak Gravity Conjecture (WGC) \cite{ArkaniHamed:2006dz}. The WGC is a well-tested Swampland constraint, and its so-called magnetic version states that in a theory with a $U(1)$ gauge symmetry, with gauge coupling $g_{U(1)}$, there is a universal quantum gravity cutoff at the scale
\begin{equation}
\label{wgc}
\Lambda_{\mathrm{WGC}} \sim g_{U(1)} M_p \;,
\end{equation}
where $M_p$ is the Planck scale. This cutoff has been argued to be associated with the appearance of an infinite number of states in a number of different ways \cite{Heidenreich:2015nta,Harlow:2015lma,Heidenreich:2016aqi,Klaewer:2016kiy,Heidenreich:2017sim,Palti:2017elp,Andriolo:2018lvp,Grimm:2018ohb,Lee:2018urn,Grimm:2018cpv}.  

Now we wish to apply the WGC to the field $\chi_{\mu}$. First we must normalise it correctly, and this is done by noting that the interaction (\ref{h1int}) should involve a quantised dimensionless charge coupling the canonically normalised field to its current. Therefore, this field should be of the form $A_{\mu} \equiv \frac{m^2 \chi_{\mu}}{M_w}$. Substituting this into the kinetic term (\ref{lfpfs}) then gives an associated gauge coupling
\begin{equation}
\label{gdef}
g_m = \frac{m}{\sqrt{2}\;M_w} \;.
\end{equation}
Now applying the WGC (\ref{wgc}) to the gauge coupling (\ref{gdef}) gives the spin-2 conjecture (\ref{s2c}). 


We note that the argument in terms of the St\"uckelberg field can also be used for massive gauge fields. This has been studied in detail in \cite{Reece:2018zvv}. An important difference however is that a gauge field can gain a mass through the Higgs mechanism. Of course, one could apply a similar argument to the longitudinal scalar mode $\pi$ in (\ref{decompo}).

\section{Bi-gravity and Relation to higher derivative terms}
\label{sec:hd}

Since the spin-2 field is massive, it is natural to consider integrating it out and obtaining an effective theory. We will consider this also for a fully non-linear background for the massive spin-2 field. 
This type of analysis has recently been studied in \cite{Gording:2018not}. One considers a bi-gravity type theory with two spin-2 fields $G$ and $W$.
\begin{equation}
\label{bigr}
\int d^4x \left[ M_p^2 \sqrt{-G} R\left(G\right) +  M_w^2\sqrt{-W} R\left(W\right)\right] \;.
\end{equation}
This theory can be supplemented by a mass term, and in case the massive field is mostly aligned with $W$, one can integrate it out. This is done on the so-called proportional background where the background values of $G$ and $W$ are proportional. The mass scale $M_w$, as defined in (\ref{h1int}), should be associated to the massive spin-2 mode. This is not quite $W$. Indeed, in general the massive mode combination of $W$ and $G$ depends on the background. However, it is possible to perform a perturbative expansion in $\frac{M_w}{M_p}$, for an appropriate background, such that the massive mode is almost aligned with $W$. In that approximation the two uses of $M_w$ agree.   

The result of integrating out the massive spin-2 mode is an infinite derivative effective action for the $G$ field, which up to quadratic order takes the form
\begin{equation}
\label{wsgw}
S_G =   \int d^4x \sqrt{-G} \left[ M_p^2 R + \frac{1}{2g_W^2}W_{\mu\nu\rho\sigma}W^{\mu\nu\rho\sigma} + \dots \right] \;,
\end{equation}
where  $W_{\mu\nu\rho\sigma}$ is the Weyl tensor
\begin{equation}
W_{\mu\nu\rho\sigma} = R_{\mu\nu\rho\sigma} + G_{\mu[\sigma}R_{\rho]\nu} + G_{\nu[\rho}R_{\sigma]\mu} + \frac13 R G_{\mu[\rho}G_{\sigma]\nu} \;. \nonumber
\end{equation}
Here $g_W$, the coefficient controlling the Weyl-squared higher derivative term, is related to the mass of the massive spin-2 field. Indeed, the pure Weyl-squared action of the form (\ref{wsgw}), without additional terms, was first studied in \cite{Stelle1978} where it was shown that it contains one massless and one massive spin-2 propagating fields. This can be seen through the double-pole propagator 
\begin{equation}
\label{prop}
\Delta_{\mu\nu\rho\sigma}(k)=\frac{1}{k^2(\frac{k^2}{2g_W^2}- M_p^2)}P_{\mu\nu\rho\sigma}\;, 
\end{equation}
where 
\begin{equation}
P_{\mu\nu\rho\sigma} = \frac12 \left( \theta_{\mu\rho}\theta_{\nu\sigma} + \theta_{\mu\sigma}\theta_{\nu\rho}\right) - \frac13 \theta_{\mu\nu} \theta_{\rho\sigma} \;, 
\end{equation}
with
\begin{equation}
\theta_{\mu\nu} = \eta_{\mu\nu} - \frac{k_{\mu}k_{\nu}}{k^2} \;.
\end{equation}
The mass of the massive spin-2 field is readily seen from (\ref{prop}) to be
\begin{equation}
\label{massw}
m = \sqrt{2}g_W M_p\;.
\end{equation}
Therefore, the spin-2 conjecture (\ref{s2c}) reads:
\begin{equation}
\label{s2cc}
\Lambda_{m} \sim \frac{g_W M_p^2}{M_w} \;.
\end{equation}
In the case $M_w=M_p$
this can be thought of as a Weak Gravity Conjecture for the Weyl-squared term. 

It should be noted that the massive spin-2 mode in the truncated version of the action (\ref{wsgw}) is a ghost. But since the original bi-gravity action does not have ghosts, one can deduce that this is an artifact of the truncation, and upon including the full infinite series of higher derivative terms the ghost becomes physical \cite{Gording:2018not}.

\section{Evidence from String Theory}
\label{sec:st}

There are many occurrences of massive spin-2 or higher fields in string theory. They can arise from string oscillator modes which fall on a Regge trajectory of increasing mass and spin. In the closed-string sector there is a tower of states, and at each level there is an additional left and right moving oscillator. The spectrum of states at each level can be deduced by decomposing the tensor product of left and right moving oscillators into representations of the Lorentz group, and always contain a spin 2 representation. In the open string sector there is only one set of oscillator modes and so the spin increases directly with the mass. At the second excited level there is a massive spin-2 field. In both cases, the massive spin-2 field is the first in an infinite tower of states with the same associated mass scale, the string scale. 

It is interesting to consider an explicit realisation of this general statement, which will also in particular illustrate the relation to higher derivative operators discussed in section \ref{sec:hd}. In \cite{Ferrara:2018wqd,Ferrara:2018wlb} a supersymmetric completion of the higher derivative Weyl-squared action (\ref{wsgw}) was constructed. The resulting theory could then be embedded in a simple string theory setting \cite{Ferrara:2018wlb}. The setting was type IIB string theory on $\mathbb{R}^{1,3} \times T^6$. A stack of $N$ space-time filling D3 brans was included, leading to a $U(N)$ gauge theory with gauge coupling $g_{D3}$ associated to the diagonal $U(1)$.\footnote{While the D3 sources an un-cancelled tadpole, the construction was primarily concerned with the representation analysis, and so issues associated to tadpole cancellation were not discussed. One could supplement the construction with mutually supersymmetric O3 planes to soak up the tadpole.} The massive spin-2 field of interest was identified as an oscillator mode in the open-string spectrum on the D3 brane. 

The mass of the massive spin-2 field can be written, in the Einstein frame, in terms of microscopic string theory parameters as $m = g_s M_s$, where $g_s$ is the string coupling and $M_s$ the string scale. This mass scale is associated with an infinite tower of open-string oscillator modes, 
\begin{equation}
m_n \sim \sqrt{n} m \;\;\forall \; n \in \mathbb{N} \;. 
\end{equation}
This infinite tower of states becomes massless in the limit $M_s\simeq (\alpha')^{-1/2}\rightarrow 0$,
and hence it matches the spin-2 conjecture. 

In string theory one naturally has $M_w=M_p$, implying that $g_m=g_W$. Then
it is interesting to note that the string theory setting leads to a relation \cite{Ferrara:2018wlb}
\begin{equation}
g_W = \frac{g_{D3}^2}{\sqrt{\cal V}}\;,
\end{equation}
where ${\cal V}$ is the volume of the internal dimensions. In the geometric regime, ${\cal V} \gg 1$, and at weak-coupling $g_{D3}\ll 1$, we have that $g_W \ll g_{D3}$. Therefore, the ultraviolet cutoff associated to the WGC is stronger when utilised as in the Spin-2 conjecture,
i.e. $\Lambda_ m\simeq g_WM_p$, than directly with the D3 brane gauge coupling $g_{D3}$, where one has $\Lambda_ {D3}\simeq g_{D3}M_p$.
From a microscopic perspective this can be understood as the statement that the coupling $g_{D3}$ refers to open-open strings coupling, while $g_W$ is an open-closed string coupling. 

Another simple way to induce massive spin-2 states is through Kaluza-Klein (KK) reduction. For simplicity we can consider a circle reduction $\mathbb{R}^{1,3} \times S^1$. Each four-dimensional field has an associated KK tower with masses, in the Einstein frame, of the form $m_n = \frac{n M_p}{\left(M_pR\right)^{\frac32}}$, where R is the circle radius. In particular, this holds for the graviton thereby inducing an infinite tower of massive spin-2 fields. In this simple setting, it is also possible to relate the St\"uckelberg gauge coupling directly to a massless $U(1)$ gauge coupling. Indeed, for the zero modes the decomposition of the 5 degrees of freedom is manifest directly as a massless graviton, graviphoton $A_R$ and dilaton $\phi_R$. The gauge coupling for the graviphoton is $g_{A} = \frac{1}{\left(M_p R\right)^{\frac32}}$. The WGC tower, with mass scale $m = g_A M_p$, is then identified with the KK tower. Note that we have also checked the analysis of section \ref{sec:s2c} holds for the case of a KK tower, where $M_w$ is found to be the Planck mass, $m$ is the KK mass, and the matter current $J^{\mu}$ is associated to KK modes of higher-dimensional matter.

Note that in \cite{Kiritsis:2006hy,Aharony:2006hz} massive spin-2 constructions were studied for AdS backgrounds holographically, and possible embeddings of these in string theory were proposed in \cite{Bachas:2017rch,Bachas:2018zmb}. It would be interesting to study if these constructions satisfy the Spin-2 conjecture.

\section{The Species Scale}
\label{sec:spsc}

While the WGC cutoff (\ref{wgc}) is associated with the appearance of an infinite tower of states, one could consider going above it by including this tower in the effective theory. It is not clear that a truncation to a finite number of elements in the tower can be consistent. In any case, there is a higher scale, the so-called Species scale \cite{Susskind:1995da,Dvali:2007hz} which forms an absolute cutoff on any effective theory due to gravity becoming strongly coupled. The species scale is defined as the scale below which there are $N$ states, so $\Lambda_S = \frac{M_p}{\sqrt{N}}$.
In terms of a KK tower, the species scale is the higher dimensional Planck scale. It also corresponds to the Landau pole for the graviphoton \cite{Harlow:2015lma,Heidenreich:2017sim,Grimm:2018ohb}. One then expects the species scale associated to the Spin-2 conjecture to satisfy
\begin{equation}
\label{spbou}
\Lambda_S < \left( \frac{m}{M_w}\right)^{\frac13} M_p\;.
\end{equation}

We can study this in the concrete string theory setups. For the circle KK reduction the relation to the species scale is as discussed above. For the open-string on D3 setting of \cite{Ferrara:2018wlb} we have both KK and string towers, and the reduction is on $T^6$ rather than a circle. In \cite{Dvali:2009ks} it was argued that the effective number of species in string theory is $N \sim \frac{1}{g_s^2}$. Then, accounting also for the KK modes of the string states, one finds a combined species scale in the Einstein frame of $\Lambda_S \sim g_s M_s$. This is the mass scale of the massive spin-2 open string mode, and so it appears that the species and spin-2 scales coincide. However, there is no contradiction because the spin-2 conjecture refers to the lightest spin-2 mass in the theory. In this case, this is actually the KK graviton with mass $m_{KK} \ll \Lambda_S$. Indeed, for an isotropic torus, counting only the KK modes, one finds an associated species scale of $\left(m_{KK}^3 M_p \right)^{\frac14}$, which satisfies the bound (\ref{spbou}). 

\section{Relation to de Sitter space}
\label{sec:ds}

The Spin-2 conjecture can be thought of as a quantum gravity obstruction to taking the massless limit for the massive spin-2 field. There is an interesting relation between this massless spin-2 limit and de Sitter space. The equations of motion that follow from the Einstein plus Weyl-squared action (\ref{wsgw}) are of the form
\begin{eqnarray}\label{weq}
  B_{\mu\nu}+{g_W^2M_p^2} G_{\mu\nu}=0,
 \end{eqnarray} 
 where   $B_{\mu\nu}$ is the Bach tensor  
\begin{eqnarray}
 B_{\mu\nu}=\nabla^\rho\nabla_\sigma W^\sigma_{~\mu\rho\nu}+
 \frac{1}{2}R^{\rho\sigma} W_{\rho\mu\sigma\nu}\, .  \label{bach}
 \end{eqnarray} 
In the massive case, for $g_W M_p\neq 0$, the equations of motion are solved by flat four-dimensional Minkowski space. 
In the massless case,  for $g_W=0$ or $M_p=0$,  first pure Weyl-squared gravity 
has flat Minkowski space as its solution. In this case the second spin-two particle becomes massless, and one also gets an additional massless vector from the $\pm 1$ helicity components of the formerly massive spin-two field.
However, the zero helicity scalar component is not physical in the massless limit on Minkowski space, since it can be gauged away by an emerging local conformal symmetry.
In the massless limit, general solutions with vanishing Bach tensor $ B_{\mu\nu}=0$
are given in terms of four-dimensional de Sitter space with arbitrary cosmological constant $\Lambda_{\mathrm{dS}}>0$.  This class of solutions is  called partially massless gravity \cite{Deser:2001pe,Hassan:2012gz,Hassan:2013pca,Garcia-Saenz:2018wnw}. Here the second spin-two field is not massless but saturates the Higuchi bound, $m^2={2\over 3} \Lambda_{\mathrm{dS}}$. Furthermore a new gauge symmetry appears in this case, which again eliminates, as in the flat case, the zero helicity component, such that the massive spin-two states has only four propagating degrees of freedom.

The possibility of allowing de Sitter solutions is in contradiction with the de Sitter Swampland Conjecture \cite{Obied:2018sgi,Ooguri:2018wrx}, as well as ideas relating to quantum break time \cite{Dvali:2013eja,Dvali:2014gua,Dvali:2017eba,Dvali:2018jhn}. This suggests a possible relation between the Spin-2 and de Sitter conjectures where they both present a quantum gravity obstruction to the same limit $g_W \rightarrow 0$.

\section{A strong Spin-2 conjecture}
\label{sec:ss2c}

The spin-2 conjecture (\ref{s2c}) is formulated for the case of Einstein gravity, with a massless graviton, and additional massive spin-2 modes. However, all the evidence presented for the conjecture holds equally if the lightest spin-2 state is actually not massless, so if the graviton itself has a mass. This is the case of (ghost-free) massive gravity \cite{deRham:2010kj}, whose action is
\begin{equation}
\int d^4x \; \sqrt{-G}\left[M_p^2 R(G) + {\cal I}\left(m,G\right) + ... \right] \;,
\end{equation}
where ${\cal I}\left(m,G\right)$ is the mass term for the graviton about a flat space background, see \cite{deRham:2014zqa} for an explicit expression.\footnote{Since we do not want to associate dynamics to the background in this setting, we are forced to consider flat-space.} 

We can then consider a stronger conjecture, which importantly applies to massive gravity theories

\medskip
\noindent
{\bf Strong Spin-2 Conjecture:} {\it An effective theory which contains a massive graviton with mass $m$, has a universal cutoff $\Lambda_{m}$ with}
\begin{equation}
\label{s2cstrong}
\Lambda_{m} \sim m\;.
\end{equation}
{\it This cutoff is associated with the mass scale of an infinite tower of states.}
 \medskip

In fact, string theory again gives further evidence for the strong spin-2 conjecture, namely in the form
  of S-fold W-superstring constructions \cite{Ferrara:2018iko}. 
  Here there is no massless spin-two graviton and
  the spectrum starts right away with massive higher spin fields: spin-two fields for open strings and spin-four fields for closed strings.
  If  the lowest massive spin-two(four) states become massless, again an entire infinite tower of higher spin states will become massless as well. 

From the Weak Gravity Conjecture perspective, the argument still holds in the sense that one can identify a St\"uckelberg gauge field, with gauge coupling $g_m = \frac{m}{\sqrt{2} M_p}$. One might be suspicious of the utilisation of the WGC in the case of a massive graviton, and this may indeed be an issue. However, there is another argument for the WGC as presented in \cite{Harlow:2015lma,Heidenreich:2017sim,Grimm:2018ohb} based on emergence of gauge symmetries. This argument is based solely on the kinetic terms and should hold irrespective of the graviton mass. 

It is worth noting that the strong Spin-2 conjecture (\ref{s2cstrong}) does not explicitly involve $M_p$. This is unusual with respect to Swampland conjectures, and this property is due to the fact that the conjecture is unique in the sense that it is about gravity itself rather than a theory coupled to gravity. The Planck mass then appears implicitly rather than explicitly.

\section{Discussion}
\label{sec:dis}

In this note we proposed a new Swampland conjecture, the Spin-2 conjecture (\ref{s2c}), which proposes a universal cutoff on any effective theory with Einstein gravity and a massive spin-2 field, that is associated to the mass scale of an infinite tower of states. In general, we related this to the Weak Gravity Conjecture, and also presented evidence from string theory. The conjecture can also be formulated in terms of higher derivative terms which lead to an additional propagating massive spin-2 mode. 

The arguments for the conjecture easily generalise to massive fields of spins larger than two. Their decomposition in terms of St\"uckelberg fields always includes a vector, to which one can apply the Weak Gravity Conjecture. 
This also matches nicely recent constraints from unitarity on higher spin theories 
\cite{Fradkin:1987ks,Vasiliev:1990en,Henneaux:2010xg,Sagnotti:2013bha}
where the mass scale of the higher spin field is associated with an infinite tower of states \cite{Afkhami-Jeddi:2018apj}.
Furthermore, as we discussed in this paper, the original spin-one Weak Gravity Conjecture implies the Spin-2 swampland conjecture. One can continue this recursively up to arbitrarily high spins. Of course, one could also work in the other direction, utilising higher spin conjectures to deduce the lower spin ones. From this perspective, one could start from the Spin-2 conjecture and deduce the Weak Gravity Conjecture. 

The Spin-2 conjecture is formulated in (\ref{s2c}) in a way which leads to the strongest constraint for the natural expectation $m \ll M_w$. However, in the inverted case $m \gg M_w$ one can simply use the Weak Gravity Conjecture with the electromagnetic dual gauge field to arrive at a strong constraint. 

The Spin-2 conjecture is most naturally applied to a theory that contains Einstein gravity, with a massless spin-2 graviton. But we expect, and formulated an appropriate strong Spin-2 conjecture (\ref{s2cstrong}), that it applies also to the case of a massive graviton. In such a setting, experiments directly constrain $m < 10^{-22}$ eV \cite{TheLIGOScientific:2016src}, while in \cite{Dvali:2006su} it was argued that a generic IR modification of gravity is constrained to have $m < 10^{-34}$ eV. The strong spin-2 conjecture would them imply the existence of an infinite tower of states at this mass scale, in sharp contradiction with observation. 

Another interesting connection with massive gravity is the relation to the strong coupling scale $\Lambda_3 = \left( m^2 M_p\right)^{\frac13}$ \cite{Deffayet:2001uk,deRham:2014zqa}. The scale $\Lambda_3$ has been shown to be closely tied to locality, unitarity and causality \cite{Bellazzini:2017fep, Bonifacio:2018aon}. For $m \ll M_p$ we see that $\Lambda_{m} \ll \Lambda_3$ and therefore the Spin-2 conjecture applies before this strong-coupling scale.

\vspace{10px}
{\bf Acknowledgements}
\noindent
We thank Marvin L\"uben and Matthew Reece for helpful discussions. We especially thank Gia Dvali and Angnis Schmidt-May for immensely illuminating and helpful explanations regarding massive and bi-gravity.
The work of D.L. is supported by the ERC Advanced Grant ``Strings and Gravity" (Grant No. 320045) and the Excellence Cluster Universe.

\bibliography{MassGraviSwamp}

\end{document}